\documentclass[11pt]{article}

\usepackage{graphicx,amsfonts,amsbsy, amssymb, amsthm, amsmath}
\usepackage[shadow]{todonotes}
\usepackage{tikz}

\usepackage{authblk}

\newcommand{\name}[1]{{#1}}

\renewcommand{\vec}[1]{{\mathbf{#1}}}

\renewcommand{\Re}{\mathop{\mathfrak{Re}}}

\newcommand{\rmd}{{\mathrm d}}
\newcommand{\rme}{{\mathrm e}}
\newcommand{\rmi}{{\mathrm i}}

\newcommand{\Ord}{{\mathrm O}}
\newcommand{\littleo}{{\mathrm o}}

\DeclareSymbolFont{lettersA}{U}{pxmia}{m}{it}

\DeclareMathSymbol{\alphaup}{\mathord}{lettersA}{"0B}
\DeclareMathSymbol{\betaup}{\mathord}{lettersA}{"0C}
\DeclareMathSymbol{\gammaup}{\mathord}{lettersA}{"0D}
\DeclareMathSymbol{\deltaup}{\mathord}{lettersA}{"0E}
\DeclareMathSymbol{\epsilonup}{\mathord}{lettersA}{"22}
\DeclareMathSymbol{\zetaup}{\mathord}{lettersA}{"10}
\DeclareMathSymbol{\etaup}{\mathord}{lettersA}{"11}
\DeclareMathSymbol{\thetaup}{\mathord}{lettersA}{"12}
\DeclareMathSymbol{\iotaup}{\mathord}{lettersA}{"13}
\DeclareMathSymbol{\kappaup}{\mathord}{lettersA}{"14}
\DeclareMathSymbol{\lambdaup}{\mathord}{lettersA}{"15}
\DeclareMathSymbol{\muup}{\mathord}{lettersA}{"16}
\DeclareMathSymbol{\nuup}{\mathord}{lettersA}{"17}
\DeclareMathSymbol{\xiup}{\mathord}{lettersA}{"18}
\DeclareMathSymbol{\piup}{\mathord}{lettersA}{"19}
\DeclareMathSymbol{\rhoup}{\mathord}{lettersA}{"1A}
\DeclareMathSymbol{\sigmaup}{\mathord}{lettersA}{"1B}
\DeclareMathSymbol{\tauup}{\mathord}{lettersA}{"1C}
\DeclareMathSymbol{\upsilonup}{\mathord}{lettersA}{"1D}
\DeclareMathSymbol{\phiup}{\mathord}{lettersA}{"1E}
\DeclareMathSymbol{\chiup}{\mathord}{lettersA}{"1F}
\DeclareMathSymbol{\psiup}{\mathord}{lettersA}{"20}
\DeclareMathSymbol{\omegaup}{\mathord}{lettersA}{"21}

\renewcommand{\Psi}{\varPsi}
\renewcommand{\Lambda}{\varLambda}
\renewcommand{\Sigma}{\varSigma}
\renewcommand{\Gamma}{\varGamma}
\renewcommand{\Theta}{\varTheta}
\renewcommand{\Xi}{\varXi}
\renewcommand{\Pi}{\varPi}
\renewcommand{\Upsilon}{\varUpsilon}
\renewcommand{\Phi}{\varPhi}
\renewcommand{\Omega}{\varOmega}

\newcommand{\R}{{\mathbb R}}
\newcommand{\N}{{\mathbb N}}
\newcommand{\Q}{{\mathbb Q}}

\newcommand{\C}{{\mathbb C}}

\newcommand{\tr}{\mathop{\mathrm{Tr}}}

\newcommand{\hash}{\#}
\newcommand{\coloneq}{\mathbin{\hbox{\raise0.08ex\hbox{\rm :}}\!\!=}}
\newcommand{\eqcolon}{\mathbin{=\!\!\hbox{\raise0.08ex\hbox{\rm :}}}}

\renewcommand{\leq}{\leqslant}
\renewcommand{\geq}{\geqslant}
\renewcommand{\epsilon}{\varepsilon} 

\newcommand{\dimostrazione}{\noindent{\sl Proof.}\phantom{X}}
\newcommand{\dimostrazionea}[1]{\noindent{\sl Proof of #1.}\phantom{X}}
\newcommand{\finire}{\hspace*{\fill}~$\Box$}

\newcommand \printdate[3]{%
    \def \@suffix##1{%
        \def \@n{##1}%
        \ifnum \@n = 1 st\else%
        \ifnum \@n = 2 nd\else%
        \ifnum \@n = 3 rd\else%
        \ifnum \@n = 21 st\else%
        \ifnum \@n = 22 nd\else%
        \ifnum \@n = 23 rd\else%
        \ifnum \@n = 31 st\else%
        th\fi \fi \fi \fi \fi \fi \fi%
    }%
    \relax%
    \number #1\raise0.7ex\hbox{\footnotesize \@suffix{#1}}\kern0.25em%
    \ifcase #2\or%
        January\or February\or March\or%
        April\or May\or June\or%
        July\or August\or September\or%
        October\or November\or December%
    \fi\ %
    \number #3%
}

\newtheorem{theorem}{Theorem}[section]
\newtheorem{proposition}[theorem]{Proposition}

\newtheorem{lemma}[theorem]{Lemma}
\newtheorem{remark}[theorem]{Remark}

\newcommand{\diag}{\mathop{\mathrm{diag}}}
\newcommand{\spam}{\mathop{\mathrm{span}}}
\newcommand{\vvec}[1]{\mathsf{#1}} 
\newcommand{\curlyC}{{\mathcal C}}
\newcommand{\curlyT}{{\mathcal T}}
\newcommand{\curlyG}{{\mathcal G}}
\newcommand{\Opf}{\mathrm{Op}(f)}

\numberwithin{equation}{section}

\usepackage[text={15cm,23cm}]{geometry} 

\usepackage[colorlinks=true,citecolor=blue,urlcolor=darkgray]
           {hyperref}  

\begin{document}
\title{Quantum ergodicity for quantum graphs without back-scattering}
\author{M.~Brammall\thanks{Present address: Department of Mathematics and 
Statistics, University of Strathclyde, Glasgow, G1 1XH, Scotland.} }
\author{B.~Winn}
\date{\printdate{8}{6}{2015}}
\affil{Department of Mathematical Sciences, 
Loughborough University, Loughborough,
LE11 3TU, U.K.}
\maketitle
\begin{abstract}
We give an estimate of the quantum variance for $d$-regular graphs
quantised with boundary scattering matrices that prohibit back-scattering.
For families of graphs that are expanders, with few short cycles, our
estimate leads to quantum ergodicity for these families of graphs. Our
proof is based on a uniform control of an associated random walk on the
bonds of the graph.  We show that recent constructions of Ramanujan graphs,
and asymptotically almost surely, random $d$-regular graphs, satisfy the
necessary conditions to conclude that quantum ergodicity holds.
\end{abstract}

\thispagestyle{empty}

\section{Introduction}
Quantum graphs have been suggested as an ideal model for problems in
quantum chaos \cite{kot:qco}.  It is therefore surprising that there
is no general theorem analogous to the Quantum Ergodicity theorem of
\v{S}nirel'man, Zelditch and Colin de Verdi\`ere
\cite{sch:epo,zel:udo,cdv:eef} in the quantum graph setting.  Quantum
ergodicity has been proved for a special class of graphs derived from
1-dimensional maps \cite{ber:qef}, and a general argument has been
presented based on physical methods of supersymmetric field theory
\cite{gnu:qeo, gnu:eso}. On the other hand it has been proved that
quantum ergodicity does not hold for quantum graphs with a star-like
configuration \cite{ber:nqe}, and entropy bounds---which control the
extent to which eigenfunctions can localise---have been derived for a
few different families of quantum graphs in \cite{kam:eof}.  A recent
article proves quantum ergodicity for the somewhat related problem of
eigenfunctions of the discrete Laplacian on combinatorial graphs
\cite{ana:qeo}.

Quantum ergodicity is one of the few universal results in quantum
chaos.  It implies a weakened form of the semi-classical eigenfunction
hypothesis \cite{ber:rai,vor:sce}, and can be stated in the following
form: let $\phi_n$ be an orthonormal basis of quantum wave-functions
with energy levels $E_n$,
$A$ an observable, with classical average $\bar{A}$. Then
\begin{equation}
  \label{eq:39}
  \lim_{E\to\infty} \frac1{\hash\{E_n\leq E\}}\sum_{E_n\leq E} \left| \langle
\phi_n, A\phi_n \rangle - \bar{A} \right|^2 = 0,
\end{equation}
provided that the classical dynamics are ergodic.  (For the reader who
prefers to keep a concrete example in mind, one can take $\phi_n$ to
be a sequence of normalised eigenfunctions of the Laplace-Beltrami
operator on a compact Riemannian manifold of negative curvature, $A$
can be a zeroth-order pseudo-differential operator with $\bar{A}$ the
mean value of its principal symbol, and the classical dynamics are the
geodesic flow on the manifold. Quantum ergodicity in this setting was
proved in \cite{sch:epo,zel:udo,cdv:eef}.)  One implication of
\eqref{eq:39} is that one can extract a density-one
subsequence\footnote{Meaning that the number of terms of the
  subsequence in a sufficiently large interval is asymptotic to the
  number of terms in the interval.} of wave-functions that becomes
equidistributed in the large energy limit.  This is equivalent to the
semi-classical eigenfunction hypothesis for that subsequence
\cite{bac:roq}.

Following the original proofs of quantum ergodicity, in the manifold
setting, the result has been proved for a variety of situations, including
Euclidian billiards \cite{zel:eoe,jak:tst}, quantised torus maps
\cite{bou:eot,mar:wla,deg:qva}, and quantised Hamiltonian flows in $\R^n$ 
\cite{hel:eel}.

In the present article we prove, deferring a precise statement of
results to the following section, a quantum ergodicity theorem for 
quantum graphs quantised with the non-back-scattering boundary 
conditions introduced in \cite{har:qgw},  provided that the underlying
graphs are expanders \cite{hoo:ega} and have not too many short cycles.
These conditions are similar to those demanded for the proof of quantum
ergodicity for eigenfunctions of the discrete Laplacian on combinatorial
graphs in \cite{ana:qeo},  although the method of proof there
is quite different.

\section{Notation and statement of results}

In order to fix notations we briefly describe the main definitions
that we will use in this work. For further background information on quantum
graphs we refer the reader to the recent book \cite{ber:itq}.

\subsection{Quantum graphs}

A quantum graph is a metric graph equipped with a differential operator
acting in a space of functions defined on the bonds of the graph. We will
denote by $\mathfrak{V}$ and $\mathfrak{B}$ respectively the set of
vertices and bonds of the graph, with $|\mathfrak{V}|=n$ and 
$|\mathfrak{B}|=B$.
The vector of bond lengths will be denoted $\vec{L}=(L_b)_{b\in\mathfrak{B}}$
where each $L_b>0$.   For us, all graphs will be undirected and 
simple, which means that
no multiple edges are allowed, nor are loops connecting vertices to
themselves.  Furthermore, we will avoid considering bipartite graphs.  This
means that the vertex set cannot be partitioned into two sets with no
connections within those sets.

Our focus will be on $d$-regular graphs, $d\geq4$, which are graphs where each
vertex is connected to $d$ other vertices. This imposes the (trivial)
constraint
\begin{equation}
  \label{eq:1}
  2B=nd.
\end{equation}

Identifying each bond $b$ of a graph $\mathcal{G}$ with an interval
$[0,L_b]$ we can define spaces $L_2(\mathcal{G})$ as the direct
product of interval $L^2$-spaces.  We will consider metric graphs
acted on by the one-dimensional (positive) Laplace operator on intervals. The
associated eigenvalue problem reads
\begin{equation}
  \label{eq:2}
  -\frac{\rmd^2}{\rmd x^2}\psi_b = k^2 \psi_b,  \qquad b\in\mathfrak{B},
\end{equation}
and solutions are bond-wise waves, with vertices as scattering points.

A vertex scattering matrix for a vertex $v$ of degree $d$ is a $d\times d$
unitary matrix $\sigma_v$, where the vector of amplitudes of incoming waves
$\vec{a}^{\mathrm{in}}\in\C^d$ is related to the vector $\vec{a}^{\mathrm{out}}$
of outgoing amplitudes by
\begin{equation}
  \label{eq:3}
  \vec{a}^{\mathrm{out}} = \sigma_v \vec{a}^{\mathrm{in}}.
\end{equation}
By considering the $2B$ \emph{directed} bonds the bond-scattering matrix
$S$ is a $2B\times 2B$ matrix with $bc$th entry equal to $0$ if
bond $b$ does not feed into a vertex $v$ that bond $c$ leaves, and otherwise
is the corresponding element of the matrix $\sigma_v$.  The matrix $S$ 
so-constructed is unitary, as a consequence of the unitarity of the $\sigma_v$.
The \emph{quantum evolution operator} $U=U(k)$ is the $2B\times 2B$ matrix
whose $bc$th entry is
\begin{equation}
U(k)_{bc} = \rme^{\rmi k L_b} S_{bc},
\end{equation}
where $L_b$ is the length\footnote{Directed bonds have the same length as their
undirected counterparts.} associated to the bond $b$.  A common convention is 
to order the directed bonds so that bonds $b=B+1,\ldots,2B$ are the reversals,
in order, of the bonds $b=1,\ldots,B$. In that case, we can write
$U(k)$ as the matrix product
\begin{equation}
  \label{eq:4}
  U(k) = D(k)S,\qquad\qquad\mbox{where } 
D(k) = \left( \begin{array}{cc}
    \rme^{\rmi k \diag \vec{L}} & 0 \\
    0 & \rme^{\rmi k \diag \vec{L}} 
  \end{array} \right),
\end{equation}
however, we will sometimes adopt a different ordering for the directed
bonds.

We define the \emph{spectrum} of the quantum graph to be the set of
non-negative values $k$ for which the condition
\begin{equation}
  \label{eq:5}
  \det(U(k)-I_{2B})=0
\end{equation}
is satisfied.  We label the points in the spectrum as $k_m$, $m=0,1,2,\ldots$,
with multiplicity, ordered so that
\begin{equation}
  \label{eq:6}
  0\leq k_0 \leq k_1 \leq k_2 \leq \cdots
\end{equation}
Condition \eqref{eq:5} is equivalent to $U(k_m)$ having an eigenvalue $1$,
and we define $\Phi_m\in\C^{2B}$ to be the corresponding eigenvector
normalised so that $\|\Phi_m\|_{\C^{2B}}=1$, or if $k_m=k_{m+1}=\cdots$ is
a multiple root of \eqref{eq:5}, take the corresponding $\Phi_m, \Phi_{m+1},
\ldots$ to be an arbitrary orthonormal basis of the eigenspace at $1$.

If $\sigma_v$ is a unitary matrix satisfying $\sigma_v^2=I_d$, then
the procedure above is equivalent to choosing a certain self-adjoint
extension of the
Laplace operator, in the sense that the spectrum defined above is the
eigenvalue set of the self-adjoint Laplace operator, with correct multiplicity
(except possibly for the eigenvalue $0$, see \cite[section 5]{ful:itf}), and 
the components of the vectors $\Phi_m$ are the amplitudes for the wave
solutions to \eqref{eq:2}, satisfying the boundary conditions implicitly
specified by the choice of extension.  To give an example,
the boundary conditions:
\begin{itemize}
\item $\psi'_{b_i}(0) = \psi'_{b_j}(0)$ for every pair of bonds $b_i, b_j$
originating at the same vertex, and,
\item $\displaystyle \sum \psi_{b}(0)=0$, where the sum is taken over all
bonds originating at a vertex,
\end{itemize}
leads, for a vertex of degree $d$, to the vertex scattering matrix
$\sigma_v$, where
\begin{displaymath}
  (\sigma_v)_{ij} = \frac{2}{d}-\delta_{ij}.
\end{displaymath}

An alternative point of view treats vertices as scattering centres, where
unitarity is a necessary condition on the matrix of transition amplitudes
to ensure probability conservation.  This point of view legitimises 
the assignment of arbitrary unitary matrices to vertices \cite{sch:ssf,tan:usm},
which allows for greater flexibility by choosing matrices with 
advantageous properties.  In \cite{har:qgw} a class of scattering matrices
were introduced with the properties that all diagonal entries are $0$, and
all off-diagonal entries have equal complex amplitude.  These were 
referred to as \emph{equi-transmitting} matrices.

The diagonal elements of a scattering matrix give the reflection amplitude
for a wave to be back-scattered into the reversal of the original bond.
By setting these elements to zero, back-scattering is prohibited in the
corresponding quantum graphs.  In \cite{har:qgw} it was speculated that
these equi-transmitting quantum graphs would lead to new advances 
in the study of quantum chaos on graphs.  Our proof in the present
article of a quantum ergodicity theorem for quantum graphs with 
equi-transmitting boundary conditions can be considered as such an
example.

Equi-transmitting matrices of size $d\times d$ have been proved to
exist in \cite{har:qgw} for $d=2^n$, $d=P+1$ where $P$ is an odd prime number,
and any $d$ for which a skew-Hadamard matrix\footnote{A skew-Hadamard
matrix $H$ is a matrix whose entries are $\pm1$, with orthogonal 
columns, and satisfying $H+H^{\mathrm{T}}=2I_d$.} exists. Equi-transmitting
matrices are further known to exist for $d=P^n+1$ for any $n>1$ and $P$ any
prime \cite{gir:private}.

\subsection{Notions of graph theory}  \label{sec:notions}
We will appeal to a few notions of graph theory, collected here for reference.
Classical graph theory is concerned with combinatorial graphs, i.e.\ without
reference to any bond lengths.  The connections are encoded by a
$n\times n$ matrix $C$ called the \emph{connectivity matrix}, whose
$ij$th entry is $1$ if vertices $i$ and $j$ are connected, and $0$ otherwise.
If the graph is without multiple edges or loops, and $d$-regular, then
$C$ is symmetric, and each row contains precisely $d$ $1$'s.

The (combinatorial) spectrum of a graph can be defined in a few ways, which
are coincident if the graph is regular. We shall define it as the set of
eigenvalues $\mu_1,\ldots,\mu_n$ of $C$.  As $C$ is symmetric, the eigenvalues
are real, and we order them in decreasing order so that
\begin{equation}
  \label{eq:7}
  -d \leq \mu_n \leq \cdots \leq \mu_1 = d.
\end{equation}
The multiplicity of the eigenvalue $d$ is the number of connected components
of the graph (so that $\mu_1=d$---every graph has at least one component),
and $\mu_n=-d$ if and only if the graph is bipartite.  The eigenvalues
of $C$ excluding $\pm d$ will be called the \emph{non-trivial} spectrum.

We will consider sequences of graphs indexed by an increasing number
of vertices $n\to\infty$.  Such a sequence of $d$-regular graphs is
called a family of expanders \cite{hoo:ega} if there exists a constant
$\beta>0$ such that the non-trivial spectrum of each graph in the
sequence is contained in the interval $[-d+\beta, d-\beta]$.  If we
can take $\beta=d-2\sqrt{d-1}$ the graphs are called
\emph{Ramanujan}. Ramanujan graphs are extremal in this sense, since
for an increasing sequence of graphs the Alon-Boppana bound
\cite[Theorem 5.3]{alo:eae,hoo:ega} implies
\begin{equation}
  \label{eq:9}
  \liminf_{n\to\infty} \mu_{2} \geq 2\sqrt{d-1}.
\end{equation}

We shall refer to \emph{cycles} on a graph, which are closed paths
without back-tracking.  We define the set $\curlyC_{B,t}$ to be the
set of bonds $b\in\mathfrak{B}$ of a graph that lie on a cycle of length 
at most $t$. 
The \emph{girth} of a graph is the length of the shortest cycle. 
In particular, this means that $\curlyC_{B,t}=\emptyset$
whenever $t$ is less than the girth.

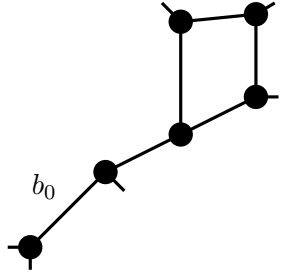
\begin{figure}
  \centering
  \begin{tikzpicture}[very thick,auto]
    \draw (-0.3,0) -- (0,0) -- (1,1) node[midway]{$b_0$} -- (2,1.5) --
    (3,2) -- (3,3.1) -- (2,3) -- (2,1.5);
    \filldraw (0,0) circle (4pt);
    \filldraw (1,1) circle (4pt);
    \filldraw (2,1.5) circle (4pt);
    \filldraw (3,2) circle (4pt);
    \filldraw (2,3) circle (4pt);
    \filldraw (3,3.1) circle (4pt);
    \draw (0,0) -- (0,-0.3);
    \draw (1,1) -- (1.25,0.75);
    \draw (3,2) -- (3.3,2);
    \draw (2,3) -- (1.75,3.25);
    \draw (3,3.1) -- (3.25,3.25);
  \end{tikzpicture}
  \caption{An illustrative example for the definition of $\curlyT_{B,t}$.  
The bond $b_0$ belongs to $\curlyT_{B,t}$ for each $t\geq 4$.}
  \label{fig:curlyT}
\end{figure}

We also define the set $\curlyT_{B,t}$ to be the set of directed bonds $b_0$
such that there exists $t_1, t_2$ with $t_1+t_2=t$ and bond $b_0$ is
a distance at most $t_1$ from a cycle of length at most $2t_2$ (see figure 
\ref{fig:curlyT}).

The sets $\curlyT_{B,t}$ and $\curlyC_{B,t}$ both give a measure of the 
number of short cycles. The set $\curlyT_{B,t}$ is more useful for our 
purposes, but $\curlyC_{B,t}$ is easier to understand.  Fortunately, their
sizes are related, as the following lemma (which does not give the
sharpest possible statement) makes clear.
\begin{lemma}
  \label{lem:TODO}
  Consider the sets $\curlyC_{B,t}$ and $\curlyT_{B,t}$ defined above, for
a $d$-regular graph.  Then
\begin{equation}
  \label{eq:35}
  | \curlyT_{B,t} | \leq \frac{(d-1)^{t-1}}{d-2} |\curlyC_{B,2t}|,
\end{equation}
where $|\cdot|$ denotes the number of elements of a set.
\end{lemma}
\dimostrazione
Clearly,
\begin{equation}
  \label{eq:36}
  |\curlyT_{B,t}| \leq \sum_{t_1+t_2=t} (d-1)^{t_1}|\curlyC_{B,2t_2}|.
\end{equation}
Furthermore, $t_2\geq 2$ for
graphs without loops or multiple 
edges, so we have
\begin{align}
  |\curlyT_{B,t}| &\leq \sum_{t_2=2}^t (d-1)^{t-t_2}|\curlyC_{B,2t_2}| \nonumber
\\
  &\leq |\curlyC_{B,2t}| \sum_{t_2=2}^t (d-1)^{t-t_2},
  \label{eq:37}
\end{align}
since $|\curlyC_{B,2t_2}| \leq |\curlyC_{B,2t}|$.  We sum the geometric 
series to get
\begin{align}
  |\curlyT_{B,t}|  &\leq
\frac{|\curlyC_{B,2t}|}{d-2} \left( (d-1)^{t-1} - 1 \right) \nonumber \\
& \leq \frac{(d-1)^{t-1}}{d-2}|\curlyC_{B,2t}|.
  \label{eq:38}
\end{align}
\finire

\subsection{Quantum Ergodicity for quantum graphs}

Observables on a quantum graph will be functions that
are constant on directed bonds, which can be represented
by members of $\C^{2B}$.  For such an observable $f\in\C^{2B}$, the
quantisation of $f$, denoted $\Opf$ is simply the diagonal 
$2B\times 2B$ matrix containing the entries of $f$:
\begin{equation}
  \label{eq:68}
  \Opf \coloneq \diag\{f\}.
\end{equation}

Let $\{\phi_j(k)\}_{j=1}^{2B}$ be an orthonormal basis of eigenvectors
of the matrix $U=U(k)$.
We define the quantum variance as, %
\begin{equation}
  \label{eq:17}
  V(f,B)\coloneq \frac1{2B}\lim_{K\to\infty}\frac1K \int_0^K 
\sum_{j=1}^{2B}
\left| \langle \phi_j(k), \Opf\phi_j(k) \rangle_{\C^{2B}} 
- {\textstyle\frac1{2B}}
\tr \Opf  \right|^2\,\rmd k.
\end{equation}
While $V(f,B)$ defined by \eqref{eq:17} does not seem immediately
analogous to \eqref{eq:39}, it was proved in \cite{ber:rbs} that
averaging the second moment appearing in \eqref{eq:17},
involving the eigenvectors of the matrix $U(k)$, over
a large window of $k$ values is equivalent to averaging an expression
similar to \eqref{eq:39} involving the eigenstates
$\Phi_0, \Phi_1, \Phi_2,\ldots$, at least for graphs with
incommensurate bond lengths.  We take \eqref{eq:17} as the starting
point of our investigation.

It turns out (see \cite{ber:qef} for example) that one cannot expect
that $V(f,B)=0$ for any individual fixed graph.  
For a fixed graph with Kirchhoff boundary conditions, a complete classification
of limiting measures induced by subsequences of eigenfunctions
has recently been obtained in \cite{cdv:scm}.  Consequently, we consider a
\emph{family} of graphs, indexed by a sequence of increasing number of bonds
$B\to\infty$.  Quantum ergodicity (sometimes called asymptotic quantum
ergodicity) for quantum graphs means that $V(f,B)\to 0$ as
$B\to\infty$ along this sequence, \cite{ber:qef,gnu:eso}.

Our main result, stated below, estimates the size of the quantum
variance for graphs quantised with equi-transmitting matrices, in
terms of certain graph-theoretic properties.

\begin{theorem}
  \label{thm:main}  For $d > 3$
  consider a $d$-regular connected, non-bipartite, simple graph, on
  $B$ bonds, quantised with equi-transmitting scattering matrices.
  Let $f\in\C^{2B}$ be an observable, satisfying $|f_b|<\kappa$ for
  all $b$, for some $\kappa>0$.  Let $T>0$, and suppose that
all non-trivial eigenvalues $\mu_i$ of the connectivity matrix of the graph 
satisfy $|\mu_i|\leq d-\beta$ for some $\beta>0$.
Then the quantum variance \eqref{eq:17} satisfies the main estimate,
\begin{equation}
  \label{eq:main}
  V(f,B) = \Ord\!\left( \frac{\kappa^2}{T\beta^2} \right) + \Ord\!\left(
\frac{ \kappa^2 (d-1)^{T} |\curlyC_{B,2T}| }{BT^2} \right),
\end{equation}
where $\curlyC_{B,2T}$ denotes the number of cycles as was described in 
section \ref{sec:notions} above.
\end{theorem}
It is natural to impose a uniform boundedness condition on observables
for which we wish to prove that quantum ergodicity to hold. This means
that the parameter $\kappa$ in the statement of theorem \ref{thm:main} 
is an absolute constant, independent of $B$.  However, the estimate
\eqref{eq:main} makes it clear how to consider observables that
grow mildly as $B\to\infty$.

Theorem \ref{thm:main} proves quantum ergodicity for families
of graphs for which a parameter $T\to\infty$ can be found as $B\to\infty$,
in such a way that the quantities on the right-hand side of \eqref{eq:main}
become negligible.  We give two examples for which this is the case:
the families of Ramanujan graphs constructed in \cite{mor:eae}, and
random $d$-regular graphs \cite{wor:mor, bol:rg1}.  We discuss these
examples further in section \ref{sec:sechs} below.

We also mention that theorem \ref{thm:main} does not require that the
bond lengths of the graph are linearly independent over $\Q$, in
contrast to many other results in this field, although if the bond
lengths are incommensurate, some values of constants can be improved---see
the comment at the end of subsection \ref{sec:ramanujan}.

We remark that we excluded bipartite graphs from our consideration in
the introduction.  Our methods could be extended to include bipartite graphs,
but there the notions of ergodicity would need to be generalised, since
bipartite graphs can support non-uniform invariant states.

All results in the present paper would hold in the case $d=3$, were it
not for the fact that no equi-transmitting matrices of size $3\times 3$
can exist (as a short calculation shows). Therefore theorem \ref{thm:main} 
is stated with condition $d>3$, although in later parts of this work
$d$ can be set equal to $3$.

In most quantum ergodicity results, such as \cite{sch:epo,zel:udo,cdv:eef},
the proof naturally separates into a semi-classical part, and a dynamical
part. In the semi-classical part, a correspondence is established between
quantum and classical time evolution, and in the dynamical part, ergodic
properties of the classical dynamics are invoked to prove quantum
ergodicity.  We shall present our proof of theorem \ref{thm:main} in this
manner. In section \ref{sec:drei} below, we relate (see proposition
\ref{prop:correspondence}) the quantum variance to a classical random walk on 
the bonds of the graph. In section \ref{sec:vier} below, we analyse the
equidistribution of the random walk, proving a uniform decay estimate
that allows us to obtain \eqref{eq:main}. The final steps of the
proof are carried out in section \ref{sec:fuenf}.

\section{Semi-classical argument}  \label{sec:drei}

Our aim is to prove, for a suitable class of $f$, that $V(f,B)\to 0$
as $B\to\infty$.  Without loss of generality we can and will assume that 
$f$ is chosen so that $\tr \Opf = 0$,
which will simplify the notation somewhat.

Our main estimate for the quantum variance goes back to an idea
from \cite{sch:otr} (see also \cite{deg:qva} for a similar application
of this idea).  Let $T>0$ and 
\begin{equation}
  \label{eq:18}
  \hat{w}_T(t) \coloneq \left\{
    \begin{array}{ll}
      \frac1{T} \left(  1-\frac{|t|}T \right) , & |t|<T, \\
      0, & \mbox{otherwise.} 
    \end{array}\right.
\end{equation}
Then for any $2B\times 2B$ unitary matrix $U$, and any orthonormal
basis $\{ \phi_j \}_{j=1}^{2B}$ of eigenvectors of $U$,
\begin{equation}
  \label{eq:19}
  \frac1{2B} \sum_{j=1}^{2B} \left| \langle \phi_j, \Opf\phi_j \rangle_{\C^{2B}}
 \right|^2 \leq \frac1{2B} \sum_{t=-T}^T \hat{w}_T(t) 
\tr(\Opf^* U^t \Opf U^{-t}).
\end{equation}
(In order to make this paper as self-contained as possible, we include
a proof of \eqref{eq:19} in an appendix.  See lemma \ref{lem:A}).  
Therefore, we can estimate the quantum variance by 
\begin{equation}
  \label{eq:8}
  V(f,B) \leq \frac1{2B} \sum_{t=-T}^T \hat{w}_T(t)\Big(
\lim_{K\to\infty} \frac1K\int_0^K \tr(\Opf^* U(k)^t \Opf U(k)^{-t})\,
\rmd k\Big).
\end{equation}
The limit in \eqref{eq:8} exists as the integrand is an almost-periodic 
function of $k$.

Let us define the matrix $\tilde{M}^{(t)}$ for $t\in\N_0$ as the 
$2B\times 2B$ matrix whose $bc$th entry is
\begin{equation}
  \label{eq:20}
  \left(\tilde{M}^{(t)}  \right)_{bc} = \lim_{K\to\infty}\frac1K
\int_0^K |U(k)^t_{bc}|^2\,\rmd k,
\end{equation}
i.e.\ the average of the square of the $bc$th element of the matrix
$U(k)^t$. Then we can rewrite the quantity inside the brackets in \eqref{eq:8}:
\begin{lemma} \label{lem:zwei}
  Let $\Opf=\diag\{f\}$ where $f\in \C^{2B}$ 
and let $\tilde{M}^{(t)}$ be defined as above. Then
\begin{equation}
 \lim_{K\to\infty}\frac1K \int_0^K  \tr(\Opf^* U(k)^t \Opf U(k)^{-t}) \,\rmd k
 = \langle f, \tilde{M}^{(t)}f \rangle_{\C^{2B}},
\end{equation}
for $t\geq 0$.
\end{lemma}
\dimostrazione 
We denote $F=\Opf$.  Expanding the trace, we have
\begin{equation}
  \label{eq:21}
  \tr(F^* U^t  F U^{-t}) = \sum_{\substack{b_0,\ldots,b_{t-1}\\
c_0,\ldots,c_{t-1}}} \bar{F}_{b_0b_0} U_{b_0b_1} U_{b_1b_2}\cdots U_{b_{t-1}c_0}
F_{c_0c_0} U_{c_0c_1}^{-1} U_{c_1c_2}^{-1} \cdots U_{c_{t-1}b_0}^{-1},
\end{equation}
where the multi-sum runs over all possible choices of $2B$ bonds for
each of $b_0,\ldots,b_{t-1}$ and $c_0,\ldots,c_{t-1}$.
We have also the expansions
\begin{equation}
  \label{eq:22}
  U^t_{b_0c_0} = \sum_{b_1,\ldots,b_{t-1}} U_{b_0b_1} U_{b_1b_2} \cdots
 U_{b_{t-1}c_0},
\end{equation}
and
\begin{align}
 \overline{U}_{b_0c_0}^t  &= \sum_{c_1,\ldots,c_{t-1}} 
\overline{U_{b_0c_{t-1}}} \overline{U_{c_{t-1}c_{t-2}}} \cdots 
\overline{U_{c_1c_0}} 
\nonumber \\
&=  \sum_{c_1,\ldots,c_{t-1}} 
 U_{c_0c_1}^{-1} U_{c_1c_2}^{-1} \cdots U_{c_{t-1}b_0}^{-1}
  \label{eq:23},
\end{align}
so that
\begin{align}
 \lim_{K\to\infty}\frac1K \int_0^K  \tr(F^* U(k)^t F U(k)^{-t}) \,\rmd k
& = \sum_{b_0, c_0} \bar{F}_{b_0b_0} \Big(
 \lim_{K\to\infty}\frac1K \int_0^K |U(k)^t_{b_0c_0}|^2 \,\rmd k \Big)
F_{c_0c_0} \nonumber \\
& = \langle f, \tilde{M}^{(t)}f \rangle_{\C^{2B}},
  \label{eq:24}
\end{align}
\finire

\begin{figure}
  \centering
  \begin{tikzpicture}[very thick,auto]
    \draw (0,0) -- (1,1) node[midway]{$b_0$}; \filldraw (1,1) circle (4pt);
    \draw (1,1) -- (1,2) node[midway]{$b_1$}; \filldraw (1,2) circle (4pt);
    \draw (1,2) -- (2,3) node[midway]{$b_2$}; \filldraw (2,3) circle (4pt);
    \draw (2,3) -- (3,3) node[midway,above]{$b_3$};  
          \filldraw (3,3) circle (4pt);
    \draw (3,3) -- (4,4) node[midway,swap]{$c_0$};
    \draw (3,3) -- (3.2,2.2) node[pos=0.7]{$c_1$};
          \filldraw (3.2,2.2) circle (4pt);
    \draw (3.2,2.2) -- (2.2,1.2) node[midway]{$c_2$};
          \filldraw (2.2,1.2) circle (4pt);
    \draw (2.2,1.2) -- (1,1) node[pos=0.6]{$c_3$};
  \end{tikzpicture}
  \caption{An example of a path and its return with $t=4$. In this example
the return path is different to the outward path. If the graph does not
contain too many cycles of length at most $6$ then this happens only
rarely.}
  \label{fig:path}
\end{figure}
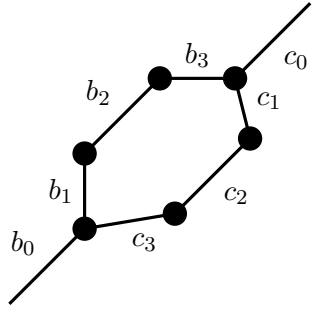

The matrix $\tilde{M}^{(t)}$ is not easy to work with, so we introduce
a second matrix $M$ where
\begin{equation}
  \label{eq:25}
  M_{bc} \coloneq  |U_{bc}|^2.
\end{equation}
Because $U$ is a unitary matrix, the matrices $M$ and $\tilde{M}^{(t)}$ are
both doubly stochastic.  If the graph does not contain too many cycles,
then the matrix $\tilde{M}^{(t)}$ is close to $M^t$, in the following sense:

\begin{proposition}  \label{prop:M}
  Let $T\in\N$ and suppose that $f\in\C^{2B}$ satisfies the bound
$|f_b|\leq \kappa$, $b=1,\ldots,2B$ for some $\kappa>0$. Provided that the
graph is quantised with scattering matrices that prohibit back-scattering,
then
\begin{equation}
  \label{eq:26}
  \left| \langle f , \tilde{M}^{(t)} f \rangle_{\C^{2B}} -
  \langle f, M^t f \rangle_{\C^{2B}}\right| \leq \frac{2\kappa^2}{(d-2)(d-1)}
(d-1)^t|\curlyC_{B,2T}|,
\end{equation}
for all $t=1,2,\ldots,T$.
\end{proposition}
\dimostrazione
By \eqref{eq:4}, we can expand $|U(k)^t_{b_0c_0}|^2$ as
\begin{equation}
  \label{eq:27}
\begin{split}
  |U(k)^t_{b_0c_0}|^2 = \sum_{\substack{b_1,\ldots,b_{t-1}\\c_1,\ldots,c_{t-1}}}
S_{b_0b_1}S_{b_1b_2}\cdots S_{b_{t-1}c_0} \rme^{\rmi k (L_{b_1}+L_{b_2}+\cdots
+L_{b_{t-1}}+L_{c_0})} \\
\times \overline{S}_{c_1c_0}\overline{S}_{c_2c_1} \cdots 
\overline{S}_{b_0c_{t-1}}
\rme^{-\rmi k(L_{c_0}+L_{c_1}+\cdots+L_{c_{t-1}})}.
\end{split}
\end{equation}
Because $S_{bc}=0$ if directed bonds $b$ and $c$ are not connected, we
can think of the right-hand side of equation \eqref{eq:27} as a
weighted sum over paths connecting $b_0$ to $c_0$ and a return path
(see figure \ref{fig:path}).

\begin{figure}
  \centering
  \begin{tikzpicture}[very thick, auto]
    \draw (0,0) -- (1,-0.5) node[midway,below]{$b_0$} -- (2,0) -- (3,-0.5) 
    -- (4,0) node[midway,below]{$c_0$};
    \filldraw (0,0) circle (4pt);
    \filldraw (1,-0.5) circle (4pt);
    \filldraw (2,0) circle (4pt);
    \filldraw (3,-0.5) circle (4pt);
    \filldraw (4,0) circle (4pt);
    \draw (2,-1) -- (2,-0) -- (2,1) -- (2.5,1.75);
    \filldraw (2,-1) circle (4pt);
    \filldraw (2,1) circle (4pt);
    \filldraw (2.5,1.75) circle (4pt);
    \node at (-1,1) {(a)};
    \draw (6,0) -- (7,-0.5) node[midway,below]{$b_0$} -- (8,0) -- (9,-0.5) 
    -- (10,0) node[midway,below]{$c_0$};
    \filldraw (6,0) circle (4pt);
    \filldraw (7,-0.5) circle (4pt);
    \filldraw (8,0) circle (4pt);
    \filldraw (9,-0.5) circle (4pt);
    \filldraw (10,0) circle (4pt);
    \draw (8,1) -- (7.5,1.5) -- (8.5,1.5) -- (8,1) -- (8,0) -- (8,-1)
    -- (9,-1) -- (8.5,-1.75) -- (8,-1);
    \filldraw (8,1) circle (4pt);
    \filldraw (7.5,1.5) circle (4pt);
    \filldraw (8.5,1.5) circle (4pt);
    \filldraw (8,-1) circle (4pt);
    \filldraw (9,-1) circle (4pt);
    \filldraw (8.5,-1.75) circle (4pt);
    \node at (5,1) {(b)};
  \end{tikzpicture}
  \caption{Two different possibilities for excursions along a path
    from $b_0$ to $c_0$. In (a) there is back-scattering; in (b) no
    back-scattering occurs but we
    note that the bond $b_0$ belongs to the set $\curlyT_{B,5}$.}
  \label{fig:excursions}
\end{figure}
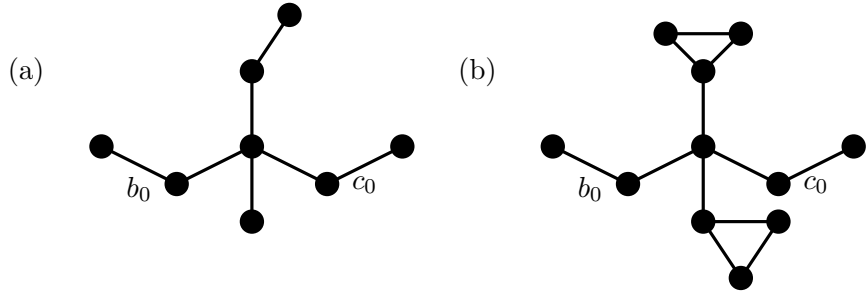

The crucial step in our argument is to demonstrate that, with few exceptions,
the return path goes back over the same bonds \emph{in the reverse order} as
the outward path.  This might not happen if, along the path, there are 
places where at least two distinct excursions from the same
vertex are made, in the sense that removing the excursions gives a shorter path
from $b_0$ to $c_0$ (see figure \ref{fig:excursions}).  Since the graphs
are quantised without back-scattering, the excursions may not consist
of self-retracing sections, so the only possibility is if the
excursions contain short cycles.

A second mechanism whereby the return path may differ from the outward
path is if the path contains a short cycle of an even number of steps
where the outward path takes one route, and the return path takes a different
route, as could happen in the situation depicted in figure \ref{fig:path}.

If the bond $b_0$ does not belong to the set $\curlyT_{B,t}$, then there
are no cycles close enough to $b_0$ to allow either possibility.  In
this case, the return path has to be the reversal of the outward path.
This means that
$c_1=b_{t-1}$, $c_2=b_{t-2}, \ldots, c_{t-1}=b_1$ and so
\begin{align}
  |U(k)^t_{b_0c_0}|^2 &= \sum_{b_1,\ldots,b_{t-1}} |S_{b_0b_1}|^2 
|S_{b_1b_2}|^2 \cdots |S_{b_{t-1}c_0}|^2 \nonumber \\
&=(M^t)_{b_0c_0},
  \label{eq:28}
\end{align}
since $|S_{bc}|=|U_{bc}|$. Since \eqref{eq:28} is independent of $k$,
the average in $k$ in \eqref{eq:20} has no effect, and we have
\begin{equation}
  \label{eq:29}
  \left( \tilde{M}^{(t)} \right)_{b_0c_0} = (M^t)_{b_0c_0},
\end{equation}
for $b_0 \notin \curlyT_{B,t}$.  Therefore, let us define a matrix $R^{(t)}$ by
\begin{equation}
  \label{eq:30}
  R^{(t)} = \tilde{M}^{(t)} - M^t,
\end{equation}
and let us consider $(R^{(t)}f)_b$, the $b$th component of $R^{(t)}f$.  We
have proved that 
\begin{equation}
  \label{eq:31}
  (R^{(t)}f)_b = 0,
\end{equation}
if $b\notin\curlyT_{B,t}$.  If $b\in\curlyT_{B,t}$ then we can be sure that
\begin{equation}
  \label{eq:32}
  |(R^{(t)}f)_b| \leq 2\kappa,
\end{equation}
since the matrices $\tilde{M}^{(t)}$ and $M$ (and hence $M^t$) are 
doubly stochastic.  Hence
\begin{align}
  \left|  \langle f , \tilde{M}^{(t)} f \rangle_{\C^{2B}} -
  \langle f, M^t f \rangle_{\C^{2B}} \right| &= \left| \langle f, R^{(t)}f
\rangle_{\C^{2B}} \right|   \nonumber \\
&\leq 2\kappa^2 |\curlyT_{B,t} | \nonumber \\
&\leq \frac{2\kappa^2}{d-2} (d-1)^{t-1} |\curlyC_{B,2t}|,
  \label{eq:33}
\intertext{using lemma \ref{lem:TODO},}
&\leq \frac{2\kappa^2}{d-2} (d-1)^{t-1} |\curlyC_{B,2T}|,\end{align}
using the fact that $|\curlyC_{B,2t}| \leq |\curlyC_{B,2T}|$ for $t\leq T$.
\finire

As we shall see below, quantum ergodicity essentially follows if we can
prove that $\langle f, \tilde{M}^{(t)}f\rangle_{\C^{2B}}=\littleo(1)$ as
$t\to\infty$.  However, we are able to prove this decay for the matrix
$M^t$ only, and proposition \ref{prop:M} provides the requisite link.  We
remark that this procedure is reminiscent of the recent proof of quantum
ergodicity for ray-splitting billiards \cite{jak:tst}.  In that work
the authors consider a probabilistic random walk on families of trajectories
with the same end-points, with transition weights given to individual
trajectories, and summed over families of trajectories; the same distinction
as between our matrices  $M^t$  and $\tilde{M}^{(t)}$, with quantum
ergodicity likewise following from ergodicity of the latter class of
random walk.  In \cite{jak:tst} the multiple trajectories arise as a 
result of splitting trajectories at the interface between one-or-more
different billiard media; in our work the multiple ``trajectories'' arise due
to the connectivity of the graphs.

The main result of this section is as follows.
\begin{proposition} \label{prop:correspondence}
  Consider a $d$-regular graph on $B$ bonds, quantised without 
back-scat\-tering.
For any $T>0$, and any $f\in\C^{2B}$ satisfying $|f_b|\leq \kappa$ for
each $b$ and $\Opf=\diag\{f\}$ satisfying $\tr \Opf=0$,
\begin{equation}
  \label{eq:34}
  V(f,B) \leq \frac1{2BT}\tr \Opf^2 + \frac1{B}\sum_{t=1}^T \hat{w}_T(t)\Big(
\langle f, M^t f \rangle_{\C^{2B}} + \Ord \left( \kappa^2(d-1)^t|\curlyC_{B,2T}|
 \right)\Big).
\end{equation}
\end{proposition}
\dimostrazione
Because of cyclic invariance of trace, and the symmetry of $\hat{w}_T$, we
can write \eqref{eq:8} (extracting the $t=0$ term) as,
\begin{equation}
  \label{eq:variance}
  V(f,B) \leq \frac{\tr \Opf^2}{2BT} + 
\frac1{B}\sum_{t=1}^T \hat{w}_T(t)\Big(
\lim_{K\to\infty} \frac1K\int_0^K \tr(\Opf^* U(k)^t \Opf U(k)^{-t})\,
\rmd k\Big).
\end{equation}
We then use lemma \ref{lem:zwei} and proposition \ref{prop:M} to get
\eqref{eq:34}. \finire

To prove our quantum ergodicity result, we will need to understand the
behaviour of $M^t f$, which represents a
random walk on the bonds of the graph.  This we do in the next section.

\section{Dynamical argument}  \label{sec:vier}
\subsection{Classical dynamics on a quantum graph}
The classical analogue of the quantum evolution is the Markov process on
the directed bonds of the graph with the transfer matrix $M$, which is
doubly-stochastic \cite{kot:pot}.  

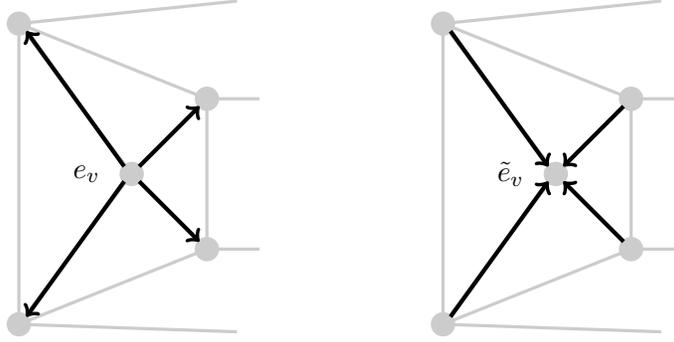
\begin{figure}[t]
  \centering
  \begin{tikzpicture}
    \begin{scope}[gray!40, very thick]
   \draw (0,0) -- (1,1) -- (1,-1) -- (0,0) -- (-1.5,2) -- (-1.5,-2) -- (0,0);
   \draw (-1.5,2) -- (1,1) -- (1.7,1);
   \draw (-1.5,-2) -- (1,-1) -- (1.7,-1);
   \draw (-1.5,2) -- (1.4,2.3);
   \draw (-1.5,-2) -- (1.4,-2.1);
   \filldraw (0,0) circle (4pt);
   \filldraw (1,1) circle (4pt);
   \filldraw (1,-1) circle (4pt);
   \filldraw (-1.5,2) circle (4pt);
   \filldraw (-1.5,-2) circle (4pt);
    \end{scope}
    \begin{scope}[->, ultra thick]
    \draw (0.1,0.1) -- (0.9,0.9);
    \draw (0.1,-0.1) -- (0.9,-0.9);
    \draw (-0.1,0.1) -- (-1.4,1.9);
    \draw (-0.1,-0.1) -- (-1.4,-1.9);
  \end{scope}
    \draw (-0.6,0) node {$e_v$};
\end{tikzpicture}
\hspace{2cm}
  \begin{tikzpicture}
    \begin{scope}[gray!40, very thick]
   \draw (0,0) -- (1,1) -- (1,-1) -- (0,0) -- (-1.5,2) -- (-1.5,-2) -- (0,0);
   \draw (-1.5,2) -- (1,1) -- (1.7,1);
   \draw (-1.5,-2) -- (1,-1) -- (1.7,-1);
   \draw (-1.5,2) -- (1.4,2.3);
   \draw (-1.5,-2) -- (1.4,-2.1);
   \filldraw (0,0) circle (4pt);
   \filldraw (1,1) circle (4pt);
   \filldraw (1,-1) circle (4pt);
   \filldraw (-1.5,2) circle (4pt);
   \filldraw (-1.5,-2) circle (4pt);
    \end{scope}
    \begin{scope}[<-, ultra thick]
    \draw (0.1,0.1) -- (0.9,0.9);
    \draw (0.1,-0.1) -- (0.9,-0.9);
    \draw (-0.1,0.1) -- (-1.4,1.9);
    \draw (-0.1,-0.1) -- (-1.4,-1.9);
  \end{scope}
    \draw (-0.6,0) node {$\tilde{e}_v$};
\end{tikzpicture}
  \caption{Support of the singular vectors $e_j$ and $\tilde{e}_j$ associated
to a vertex $v$. See main text for the definitions.}
  \label{fig:singular}
\end{figure}

Since $M$ is not necessarily normal, it will be convenient to work,
rather than with eigenvectors of $M$, with its singular vectors,
defined to be the eigenvectors of the symmetric matrix $M^TM$. If we order
the directed bonds in groups of $d$ bonds departing from each vertex,
the matrix $M^TM$ decomposes into block-diagonal form
\begin{displaymath}
  M^TM = \left( \begin{array}{cccc}
      J & 0 & \cdots & 0 \\
      0 & J & \cdots & 0 \\
      \vdots & \vdots & \ddots & \vdots \\
      0 & 0 & \cdots & J 
    \end{array} \right),
\end{displaymath}
where each of the $n$ blocks is a $d\times d$ matrix
\begin{displaymath}
  J = \frac1{(d-1)^2} \left( \begin{array}{cccc}
      d-1 & d-2 & \cdots & d-2 \\
      d-2 & d-1 & \cdots & d-2 \\
      \vdots & \vdots & \ddots & \vdots \\
      d-2 & d-2 & \cdots & d-1 
    \end{array} \right). 
\end{displaymath}
The eigenvalues of $J$ are $1$ with multiplicity one, and $(d-1)^{-2}$ with
multiplicity $d-1$, and the simple
eigenspace is spanned by the vector $(1,\ldots,1)^T$.  Therefore the
singular values of the matrix $M$ are: $1$ with multiplicity $n$, and
$(d-1)^{-1}$ with multiplicity $n(d-1)$. A basis for the eigenspace of
$M^TM$ with eigenvalue $1$ is given by the set $\{ e_1,\ldots,e_n \} 
\subseteq \C^{2B}$, where the $j^{\mathrm th}$ component of $e_v$ is
defined to be $1$ if directed bond $j$ points outwards from vertex
$v$, and $0$ otherwise (see figure~\ref{fig:singular}).

We could equally-well consider $MM^T$, which has identical spectrum to
$M^TM$, and a basis of eigenvectors corresponding to the singular
value $1$ can be chosen to have zero components except for the
\emph{incoming} bonds of vertex $v$ where the component is $1$. We
will denote these vectors by $\tilde{e}_v\in \C^{2B}$ for
$v=1,\ldots,n$ (see figure \ref{fig:singular}).

Observables that are linear combinations of $e_1,\ldots,e_n$ 
will be called \emph{evenly distributed around vertices} in the
following subsection.

\begin{proposition} \label{prop:zwei}  
If the scattering matrices on the quantum graph are equi-transmitting,
the action of $M$ on the vectors $e_v$ and $\tilde{e}_v$ is as follows:
\begin{align}
Me_v &= \tilde{e}_v, \\
M\tilde{e_v} &= \frac{1}{d-1}\left(\sum_{w \sim v}\tilde{e}_w - e_v \right),
\end{align} 
where $\sum_{w \sim v}$ is a sum over vertices $w$ connected to $v$.
\end{proposition}
\dimostrazione
The action of $M$ on a vector supported on a single directed bond $b$ is
to allocate an equal fraction to the $d-1$ directed bonds feeding into
the origin of $b$ (not including the reversal of $b$). For the vector
$Me_v$, each bond directed towards $v$ gets $d-1$ times the fraction
$1/(d-1)$ of the weight $1$ on each outward pointing bond in $e_v$. The
result is a vector of weight $1$ on each inward pointing vertex to $v$.
Hence $Me_v = \tilde{e}_v$.

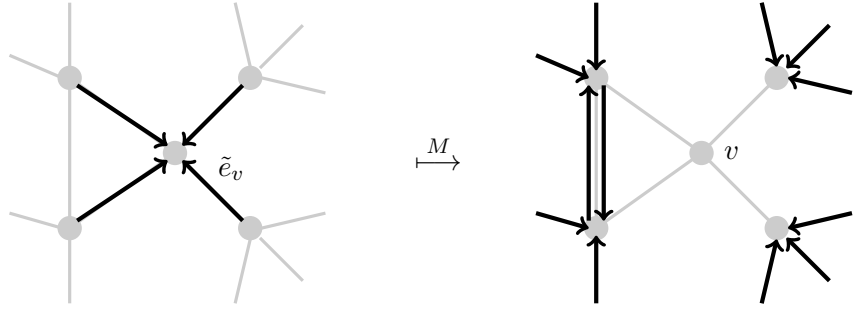
\begin{figure}[t]
  \centering
  \begin{tikzpicture}
    \begin{scope}[gray!40, very thick, shift={(7,0)}]
      \filldraw (0,0) circle (4pt); 
      \draw (0,0) -- (1,1); \filldraw (1,1) circle (4pt);
      \draw (0,0) -- (1,-1); \filldraw (1,-1) circle (4pt);
      \draw (0,0) -- (-1.4,1); \filldraw (-1.4,1) circle (4pt);
      \draw (0,0) -- (-1.4,-1); \filldraw (-1.4,-1) circle (4pt);
      \draw (-1.4,-1) -- (-1.4,1);
    \end{scope}
    \draw (7.4,0) node {$v$};
    \begin{scope}[->, ultra thick, shift={(7,0)}]
      \draw (-1.4,2) -- (-1.4,1.1);
      \draw (-2.2,1.3) -- (-1.5,1);
      \draw (-1.4,-2) -- (-1.4,-1.1);
      \draw (-2.2,-0.8) -- (-1.5,-1);
      \draw (-1.5,-0.9) -- (-1.5,0.9);
      \draw (-1.3,0.9) -- (-1.3,-0.9);
      \draw (1.7,1.7) -- (1.13,1.13);
      \draw (1.7,-1.7) -- (1.13,-1.13);
      \draw (0.8,2) -- (1,1.15);
      \draw (2,0.8) -- (1.15,1);
      \draw (0.8,-2) -- (1,-1.15);
      \draw (2,-0.8) -- (1.15,-1);
    \end{scope}
    \begin{scope}[gray!40, very thick]
      \filldraw (0,0) circle (4pt);
      \draw (0,0) -- (1,1); \filldraw (1,1) circle (4pt);
      \draw (0,0) -- (1,-1); \filldraw (1,-1) circle (4pt);
      \draw (0,0) -- (-1.4,1); \filldraw (-1.4,1) circle (4pt);
      \draw (0,0) -- (-1.4,-1); \filldraw (-1.4,-1) circle (4pt);
      \draw (-1.4,-1) -- (-1.4,1);
      \draw (-1.4,2) -- (-1.4,1.1);
      \draw (-2.2,1.3) -- (-1.5,1);
      \draw (-1.4,-2) -- (-1.4,-1.1);
      \draw (-2.2,-0.8) -- (-1.5,-1);
      \draw (1.7,1.7) -- (1.13,1.13);
      \draw (1.7,-1.7) -- (1.13,-1.13);
      \draw (0.8,2) -- (1,1.15);
      \draw (2,0.8) -- (1.15,1);
      \draw (0.8,-2) -- (1,-1.15);
      \draw (2,-0.8) -- (1.15,-1);
    \end{scope}
    \draw (0.75,-0.2) node {$\tilde{e}_v$};
    \begin{scope}[->, ultra thick]
      \draw (-1.3,-0.9) -- (-0.1,-0.1);
      \draw (-1.3,0.9) -- (-0.1,0.1);
      \draw (0.9,0.9) -- (0.1,0.1);
      \draw (0.9,-0.9) -- (0.1,-0.1);
    \end{scope}
  \draw (3.5,0) node {$\overset{M}{\longmapsto}$};
  \end{tikzpicture}
  \caption{The support of $M\tilde{e}_v$. The vertex $v$ is at the centre of the illustration.}
  \label{fig:vier}
\end{figure}

In a similar way, it is clear that $M\tilde{e}_v$ has weight $1/(d-1)$ on
each bond directed towards a neighbour of $v$, except the outward 
pointing bonds from $v$ (see figure \ref{fig:vier}). Such a vector
can be written
\begin{displaymath}
  \frac{1}{d-1}\left(\sum_{w \sim v}\tilde{e}_w - e_v \right).
\end{displaymath}
\finire

\subsection{Observables evenly distributed around vertices}
Let
\begin{equation}
  G_1\coloneq \spam\{ e_1,\ldots,e_n\} \subseteq \C^{2B},
\end{equation}
and let $\varphi:G_1\to\C^n$ be the natural isomorphism. Let us also
define $\tilde{\varphi}: G_1 \to \C^{2n}$ by,
\begin{equation}
  \tilde{\varphi}: a_1 e_1 +\cdots + a_n e_n \mapsto \left(\begin{array}{c}
a_1 \\ \vdots \\ a_n \\ 0 \\ \vdots \\ 0 \end{array}\right).  
\end{equation}

We shall also work
with the larger space $\widetilde{G_1}$, where
\begin{equation}
  \widetilde{G_1}\coloneq \spam\{e_1,\ldots,e_n,\tilde{e}_1,\ldots,\tilde{e}_n
  \} \subseteq \C^{2B},
\end{equation}
and define $\psi:\C^{2n}\to \widetilde{G_1}$ by
\begin{equation}
  \psi: \left(\begin{array}{c} \vvec{a} \\ \vvec{b} \end{array}\right) 
  \mapsto a_1 e_1 + \cdots + a_n e_n + b_1 \tilde{e}_1 + \cdots + b_n
  \tilde{b}_n,
\end{equation}
where
\begin{equation}
  \vvec{a} = \left(\begin{array}{c} a_1 \\ \vdots \\ a_n \end{array}\right).
\end{equation}
We note that $\psi$ fails to be invertible since 
\begin{equation}
  \psi\left( \begin{array}{c} \vvec{e} \\ -\vvec{e} \end{array}\right) = 0,
\qquad\text{where }\vvec{e} = \left(\begin{array}{c} 1 \\ \vdots \\ 1 
\end{array}\right).
\end{equation}

\begin{lemma} \label{lem:eins}
  If $\hat{x}=\left(\begin{array}{c} \vvec{x} \\ \vvec{y} \end{array}\right)
\in\C^{2n}$ then
\begin{equation}  \label{eq:anna}
  \| \psi(\hat{x}) \|_{\C^{2B}} \leq \sqrt{d}\left( \| \vvec{x} \|_{\C^n}
    + \| \vvec{y} \|_{\C^n}\right).
\end{equation}
\end{lemma}
\dimostrazione
Let us define
\begin{align*}
 \eta &= \sum_{j=1}^n x_j e_j \\
 \tilde\eta &= \sum_{j=1}^n y_j \tilde{e}_j,
\end{align*}
so that
\begin{equation}
  \label{eq:10}
  \| \psi(\hat{x}) \|_{\C^{2B}}^2 = \| \eta + \tilde\eta \|^2_{\C^{2B}}
=\| \eta \|^2_{\C^{2B}} + \| \tilde\eta \|^2_{\C^{2B}} + 2\Re\langle \eta,
\tilde\eta \rangle_{\C^{2B}}.
\end{equation}
Due to orthogonality of $\{e_1,\ldots,e_n\}$,
\begin{equation}
  \label{eq:11}
  \| \eta \|_{\C^{2B}}^2 = \sum_{j=1}^n |x_j|^2\| e_j \|_{\C^{2B}}^2 
  = d \sum_{j=1}^n |x_j|^2 = d\| \vvec{x} \|_{\C^n}^2,
\end{equation}
and similarly
\begin{equation}
  \label{eq:12}
  \| \tilde\eta \|_{\C^{2B}}^2 = d \| \vvec{y} \|_{\C^n}^2.
\end{equation}
We also have
\begin{equation}
  \label{eq:13}
  \langle \eta, \tilde\eta \rangle_{\C^{2B}} = \sum_{i,j=1}^n x_i\bar{y}_j
\langle e_i, \tilde{e}_j \rangle_{\C^{2B}}.
\end{equation}
Now, since $e_j$ is supported on outward pointing bonds from vertex $j$
and $\tilde{e}_i$ is supported on inward pointing bonds to vertex
$i$, the only way that $\langle e_i , \tilde{e}_j \rangle_{\C^{2B}}$ 
can be non-zero is if $i$ connects to $j$. We have, in fact,
\begin{equation}
  \label{eq:14}
  \langle e_i , \tilde{e}_j \rangle_{\C^{2B}} = \left\{
    \begin{array}{l}
      \mbox{$1$, if $i\sim j$,} \\
      \mbox{$0$, otherwise,}
    \end{array} 
\right\}  = C_{ij},
\end{equation}
so
\begin{align}
  \label{eq:15}
  \langle \eta, \tilde\eta \rangle_{\C^{2B}} &= \sum_{i,j=1}^n x_i \bar{y}_j
C_{ij}  \nonumber \\
 &=\langle \vvec{x}, C\,\vvec{y} \rangle_{\C^n}.
\end{align}
We get,
\begin{align}
  \label{eq:16}
  \| \psi(\hat{x}) \|_{\C^{2B}}^2 &= d \| \vvec{x} \|_{\C^n}^2 +
d\| \vvec{y} \|_{\C^n}^2 + 2\Re \langle \vvec{x}, C\,\vvec{y} \rangle_{\C^n} 
\nonumber \\
&\leq   d \| \vvec{x} \|_{\C^n}^2 +
d\| \vvec{y} \|_{\C^n}^2 + 2  \| \vvec{x} \|_{\C^n} 
\| C\,\vvec{y} \|_{\C^n} \nonumber \\
&\leq   d \| \vvec{x} \|_{\C^n}^2 +
d\| \vvec{y} \|_{\C^n}^2 + 2d  \| \vvec{x} \|_{\C^n} 
\| \vvec{y} \|_{\C^n} \nonumber \\
&= d(  \| \vvec{x} \|_{\C^n} +
\| \vvec{y} \|_{\C^n})^2,
\end{align}
proving \eqref{eq:anna}.
\finire

Let $\hat{C}$ be the $2n\times 2n$ matrix,
\begin{equation}
 \hat{C} \coloneq \left( \begin{array}{rr} 0 & \frac{-1}{d-1}I_n   \\
 I_n & \frac1{d-1}C \end{array}\right).
\end{equation}
\begin{proposition}
  Let $f\in G_1$ and $t=0,1,2,\ldots$ Then
  \begin{equation}  \label{eq:psi}
    \psi\left( \hat{C}^t\tilde\varphi(f)\right) = M^t f.
  \end{equation}
\end{proposition}
\dimostrazione
Let $\hat{x}\in \C^{2n}$ with $g=\psi(\hat{x})$. We first prove that
\begin{equation} \label{eq:step}
\psi\left( \hat{C}\hat{x}\right) = M g.
\end{equation}
Indeed, suppose that
\begin{equation}
  g = x_1 e_1 + \cdots + x_n e_n + \tilde{x}_1 \tilde{e}_1 + \cdots
+ \tilde{x}_n \tilde{e}_n,
\end{equation}
so that
\begin{equation}
  Mg = x_1 \tilde{e}_1 + \cdots x_n \tilde{e}_n + \frac{\tilde{x}_1}{d-1}
  \Big( \sum_{j\sim 1} \tilde{e}_j - e_1 \Big) + \cdots + 
\frac{\tilde{x}_n}{d-1}  \Big( \sum_{j\sim n} \tilde{e}_j - e_n \Big),
\end{equation}
by proposition \ref{prop:zwei}.
However, with $\vvec{x}=(x_1,\ldots, x_n)^T$, $\tilde{\vvec{x}} = 
(\tilde{x}_1,\ldots,\tilde{x}_n)^T$ and $\hat{x}=(\vvec{x}, 
\tilde{\vvec{x}})^T$, we have
\begin{equation}
  \hat{C}\hat{x} = \left(\begin{array}{c} -\frac{\tilde{\vvec{x}}}{d-1} \\
\vvec{x} + \frac{C\tilde{\vvec{x}}}{d-1}\end{array}\right),  
\end{equation}
from which we see that $\psi(\hat{C}\hat{x}) = Mg.$

To prove \eqref{eq:psi} in the case $t=0$, it is immediate to observe
that $\psi(\tilde{\varphi}(f)) = f$, from the definitions of $\psi$
and $\tilde{\varphi}$. If we assume that $\psi(\hat{C}^t \tilde{\varphi}(f))
=M^tf$, then applying \eqref{eq:step} with $\hat{x}=\hat{C}^t\tilde{\varphi}
(f)$ proves \eqref{eq:psi} for the $t+1$ case. \finire

Therefore, to understand the action of $M$ on $f$, we need to understand
the iterates of $\hat{C}$. 

Let $\vvec{x}$ be an eigenvector of $C$ with eigenvalue $\mu$. Then
\begin{equation}
  \hat{C}\left(\begin{array}{c} \vvec{x} \\ 0 \end{array}\right) =
  \left( \begin{array}{c} 0 \\ \vvec{x} \end{array}\right).
\end{equation}
Denoting 
\begin{align}
  \hat{C}^t \left(\begin{array}{c} \vvec{x} \\ 0 \end{array}\right) &\eqcolon
  \left( \begin{array}{c} y_t\vvec{x} \\ z_t \vvec{x} \end{array}\right),
\label{eq:Chat_n} \\ &=\left(
\begin{array}{c} -\frac{z_{t-1}}{d-1}\vvec{x} \\ ( y_{t-1} + \frac{\mu z_{t-1}}
{d-1} )\vvec{x} \end{array}\right).
\end{align}
So
\begin{equation}
  y_{t} = -\frac{z_{t-1}}{d-1},
\end{equation}
and $z_t = z_t(\mu)$ is the solution to the recurrence
\begin{equation}
  \label{eq:z_recurrence}
  z_t = \frac{\mu z_{t-1}}{d-1} - \frac{z_{t-2}}{d-1},
\end{equation}
with initial conditions $z_0=0$ and $z_1=1$.

\begin{proposition} \label{prop:drei}
If $|\mu|\leq d-\beta < d$, then the solutions to 
\eqref{eq:z_recurrence} satisfy
\begin{equation}
  \label{eq:z_bound}
  |z_t(\mu)| \leq t \left( \frac{d-1-\beta}{d-1} \right)^{t-1},
\end{equation}
for $t=1,2,\ldots$
\end{proposition}
\dimostrazione 
We first consider the case $|\mu|\neq 2\sqrt{d-1}$. Standard methods yield
the solution to (\ref{eq:z_recurrence}) in this case to be given by
\begin{equation} \label{eq:z_omega}
  z_t = \frac{\sqrt{d-1}}{2\sqrt{\omega^2-1}}\left(\left( \frac{\omega+
\sqrt{\omega^2-1}}{\sqrt{d-1}}\right)^t - \left(  \frac{\omega-
\sqrt{\omega^2-1}}{\sqrt{d-1}}\right)^t \right),
\end{equation}
where $\omega = \frac12\mu(d-1)^{-1/2}\neq \pm 1$. Since
\begin{equation}
 \left( \frac{\omega+
\sqrt{\omega^2-1}}{\sqrt{d-1}}\right) - \left(  \frac{\omega-
\sqrt{\omega^2-1}}{\sqrt{d-1}}\right) = 2\frac{\sqrt{\omega^2-1}}{\sqrt{d-1}},
\end{equation}
we may apply the inequality
\begin{equation}
 \left| \frac{a^t-b^t}{a-b}\right| \leq t\max\{|a|,|b|\}^{t-1},
\end{equation}
to (\ref{eq:z_omega}), to get
\begin{equation}
 |z_t|\leq t\max\left\{  \left|\frac{\omega+
\sqrt{\omega^2-1}}{\sqrt{d-1}}\right| , \left|  \frac{\omega-
\sqrt{\omega^2-1}}{\sqrt{d-1}}\right|\right\}^{t-1}.
\end{equation}
For $1<\omega\leq \frac12(d-\beta)(d-1)^{1/2}$ we have
\begin{equation}
  \left| \frac{\omega\pm\sqrt{\omega^2-1}}{\sqrt{d-1}} \right| \leq
  \frac{\omega+\sqrt{\omega^2-1}}{(d-1)^{1/2}}.
\end{equation}
For such values of $\omega$,
\begin{equation}
  \label{eq:88}
  0\leq \omega^2-1 \leq \frac{(d-\beta-2)^2-4\beta}{4(d-1)} <
  \frac{(d-\beta-2)^2}{4(d-1)},
\end{equation}
so that
\begin{align}
\nonumber \\
  \left| \frac{\omega\pm\sqrt{\omega^2-1}}{\sqrt{d-1}} \right|
  &<  \frac12\frac{d-\beta}{d-1} + \frac{d-2-\beta}{2(d-1)} \\ \nonumber
  &= \frac{d-1-\beta}{d-1}.
\end{align}
A similar argument holds if $-\frac12(d-\beta)(d-1)^{1/2}\leq \omega < -1$.

If $|\omega|<1$ then
\begin{equation}
  \left| \frac{\omega\pm\sqrt{\omega^2-1}}{\sqrt{d-1}} \right| = 
  \frac{1}{\sqrt{d-1}} \leq \frac{d-1-\beta}{d-1}.
\end{equation}

Finally, in the case $|\mu|=2\sqrt{d-1}$, directly solving 
(\ref{eq:z_recurrence}), we find
\begin{equation}
  |z_t(\mu)| = \frac{t}{(d-1)^{(t-1)/2}}.
\end{equation}
\finire

Our main result of this subsection is the following:
\begin{proposition} \label{prop:vier}
Let $f\in G_1$ with $\tr \Opf=0$. 
If all non-trivial eigenvalues $\mu$ of $C$ satisfy the bound
$|\mu|< d-\beta$ then we have
  \begin{equation}
    \| M^t f \|_{\C^{2B}} \leq 2 \| f \|_{\C^{2B}} t \left( \frac{d-1-\beta}
{d-1} \right)^{t-1},    
  \end{equation}
for $t=1,2,3,\ldots$
\end{proposition}
\dimostrazione
Let $\vvec{x}_1,\ldots,\vvec{x}_n$ be an orthonormal basis of eigenvectors
of $C$, where $\vvec{x}_1=n^{-1/2}\vvec{e}$. Since 
$\tr \Opf= 0$ we have $\langle \varphi(f), \vvec{e} \rangle_{\C^n}=0$
and we can write
\begin{equation}
  \varphi(f) = \alpha_2 \vvec{x}_2 + \cdots + \alpha_n \vvec{x}_n,  
\end{equation}
where
\begin{equation}
  \| \varphi(f) \|_{\C^n}^2 = |\alpha_2|^2 + \cdots + |\alpha_n|^2.  
\end{equation}
It is also easy to see that
\begin{equation}
 \| f \|_{\C^{2B}}^2 = d  \| \varphi(f) \|_{\C^n}^2.
\end{equation}

We have
\begin{equation}
  \tilde\varphi(f) = 
 \alpha_2 \left(\begin{array}{c} \vvec{x}_2 \\ 0 \end{array} \right)
 + \cdots + 
 \alpha_n \left(\begin{array}{c} \vvec{x}_n \\ 0 \end{array} \right) 
 \in \C^{2n}.
\end{equation}
So, if $t=1,2,3,\ldots$,
\begin{equation}
  \hat{C}^t \tilde\varphi(f) =   
 \frac{\alpha_2}{d-1} \left(\begin{array}{c} -z_{t-1}(\mu_2)\vvec{x}_2 \\
 (d-1)z_t(\mu_2)\vvec{x}_2 \end{array} \right)
 + \cdots + 
 \frac{\alpha_n}{d-1} \left(\begin{array}{c} -z_{t-1}(\mu_n)\vvec{x}_n \\
 (d-1)z_t(\mu_n)\vvec{x}_n \end{array} \right),
\end{equation}
using (\ref{eq:Chat_n}).
We now use lemma \ref{lem:eins} to get
\begin{align}
  \| M^t f \|_{\C^{2B}} &= \left\| \psi\left( \hat{C}^t \tilde\varphi(f) 
      \right)\right\|_{\C^{2B}} \nonumber \\
    &\leq \sqrt{d} \left( \Bigg( \sum_{j=2}^n |\alpha_j|^2 \frac{|z_{t-1}(
        \mu_j)|^2}{(d-1)^2} \Bigg)^{1/2} +
      \Bigg( \sum_{j=2}^n |\alpha_j|^2 |z_t(\mu_j)|^2 \Bigg)^{1/2} \right) 
    \nonumber \\
    & \leq \sqrt{d} \Bigg(\sum_{j=2}^n |\alpha_j|^2 \Bigg)^{1/2}
    \frac{(t-1)(d-1-\beta)^{t-2} + t(d-1-\beta)^{t-1}}{(d-1)^{t-1}},
\nonumber \\
\intertext{using proposition \ref{prop:drei},} 
&\leq 2 \| f \|_{\C^{2B}} t \left( \frac{d-1-\beta}{d-1} \right)^{t-1}.
\end{align}
\finire
\subsection{General mean-zero observables}

We would like to consider a more general class of observables than those
belonging to the spaces
$G_1$.  Let $G_2 = G_1^{\perp}$. Thus, $G_2$ is the space of singular 
vectors of $M$ with singular value $\frac{1}{d-1}$, and it follows that
$M$ acts on $G_2$ by contraction:

\begin{lemma}\label{lem:singular}
Let $g \in G_2$. Then,
\begin{equation}\nonumber
||Mg||_{\mathbb{C}^{2B}} = \frac{||g||_{\mathbb{C}^{2B}}}{d-1}
\end{equation}
\end{lemma}

\dimostrazione
Take $g \in G_2$. Then, we have  
\begin{align}
||Mg||^2_{\mathbb{C}^{2B}} = \langle Mg, Mg \rangle_{\mathbb{C}^{2B}} &= \langle g, M^TMg \rangle_{\mathbb{C}^{2B}} \\
&= \frac{||g||^2_{\mathbb{C}^{2B}}}{\left(d-1\right)^2}.
\end{align}
\finire

Finally, let us put the results of proposition~\ref{prop:vier} and lemma
\ref{lem:singular} together.

\begin{theorem}\label{thm:general}
  Let $f\in\C^{2B}$ with $\tr \Opf =0$. Assume that all non-trivial eigenvalues
$\mu$ of the connectivity matrix satisfy $|\mu|\leq d-\beta$ for some
$\beta>0$. Then there exists a constant $K>0$ (that does not depend
on $t$ or $B$) such that for $t\geq 1$,
\begin{equation}
  \| M^t f \|_{\C^{2B}} < K
\|f\|_{\C^{2B}} t \left( \frac{d-1-\beta}{d-1} \right)^t.
\end{equation}
\end{theorem}
\dimostrazione
The conditions of the theorem guarantee that
\begin{equation}
  \label{eq:a_bound}
    \| M^j g \|_{\C^{2B}} \leq 2 \| g \|_{\C^{2B}} j \left( \frac{d-1-\beta}
{d-1} \right)^{j-1}  
\end{equation}
for any $g\in G_1$ for $j\in\N$, due to proposition \ref{prop:vier}.

We decompose $f$ according to $\C^{2B}=G_1\oplus G_2$ as 
$f = f_{0,1} + f_{0,2}$, where $f_{0,1}\in G_1$ and $f_{0,2}\in G_2$, 
and inductively defining the sequences
$\{f_{j,1}\}_{j=0}^{\infty}$ and  $\{f_{j,2}\}_{j=0}^{\infty}$ by
\begin{equation}
Mf_{j,2} = f_{j+1,1} + f_{j+1,2},
\end{equation}
where $f_{j,i}\in G_i$ for $j\geq 0$, $i\in\{1,2\}$. So upon each iteration,
the component of $f_{j,2}$ that does not remain in $G_2$ becomes 
$f_{j+1,1}\in G_1$.

We have, by lemma \ref{lem:singular}
\begin{equation}
  \frac{\|f_{j,2}\|_{\C^{2B}}^2}{(d-1)^2} =
||Mf_{j,2}||^2_{\C^{2B}} = 
||f_{j+1,1}||^2_{\C^{2B}} + ||f_{j+1,2}||^2_{\C^{2B}} 
 \geq ||f_{j+1,i}||^2_{\C^{2B}},
\end{equation}
for $i\in\{1,2\}$, so, inductively,
\begin{equation} \label{eq:iter_1}
\| f_{j,2} \|_{\C^{2B}} \leq \frac{||f_{0,2}||_{\C^{2B}}}{\left(d-1\right)^{j}}
\leq \frac{||f||_{\C^{2B}}}{\left(d-1\right)^{j}},
\end{equation}
and 
\begin{equation} \label{eq:iter_2}
\| f_{j,1} \|_{\C^{2B}} \leq \frac{\|f_{j-1,2}\|_{\C^{2B}}}{d-1} 
\leq \frac{||f||_{\C^{2B}}}{\left(d-1\right)^{j}}.
\end{equation}

Acting on $f$ iteratively, we have
\begin{align}
M^tf &= M^tf_{0,1}+ M^tf_{0,2} \nonumber \\
&= M^tf_{0,1} + M^{t-1}f_{1,1}+M^{t-1}f_{1,2} \nonumber \\
&\;\,\vdots \nonumber \\
&= \sum_{j=0}^{t-1}M^{t-j}f_{j,1} + f_{t,1} + f_{t,2}.
\end{align}

Thus, we have, using \eqref{eq:iter_1} and \eqref{eq:iter_2}, and
\eqref{eq:a_bound},
\begin{align}
\|M^tf\|_{\C^{2B}} &\leq
 \sum_{j=0}^{t-1}\|M^{t-j}f_{j,1}\|_{\C^{2B}} + 
\|f_{t,1}\|_{\C^{2B}} + \|f_{t,2}\|_{\C^{2B}} \nonumber \\
&\leq 2\sum_{j=0}^{t-1}(t-j)\left(\frac{d-1-\beta}{d-1}\right)^{t-j-1}
||f_{j,1}||_{\C^{2B}} + \frac{2||f||_{\C^{2B}}}{\left(d-1\right)^t} \nonumber \\
&\leq 2\sum_{j=0}^{t-1}(t-j)\left(\frac{d-1-\beta}{d-1}\right)^{t-j-1}
\frac{||f||_{\C^{2B}}}{\left(d-1\right)^j} + \frac{2||f||_{\C^{2B}}}
{\left(d-1\right)^t} \nonumber \\
&=\frac{2||f||_{\C^{2B}}}{\left(d-1\right)^{t-1}}
\sum_{r=1}^{t}r(d-1-\beta)^{r-1} + \frac{2||f||_{\C^{2B}}}
{\left(d-1\right)^t} \qquad\text{\textit{via} $r=t-j$,} \nonumber \\
&\leq 2t\|f\|_{\C^{2B}}(d-1)\frac{\left(d-1-\beta\right)^t-1}
{\left(d-1\right)^t\left(d-2-
\beta\right)} + \frac{2\|f\|_{\C^{2B}}}{\left(d-1\right)^t} \nonumber \\
&\leq \frac{2t\|f\|_{\C^{2B}}(d-1)}{\left(d-2-
\beta\right)}\left(\frac{d-1-\beta}{d-1}\right)^t + 
\frac{2\|f\|_{\C^{2B}}}{\left(d-1\right)^t}.
\end{align}
Finally (to combine the two terms into a single), noting that
\begin{equation}
  \frac{t(d-1)(d-1-\beta)^t}{d-2-\beta} \geq 4,  
\end{equation}
for $t\geq 1$, we get
\begin{equation}
  \| M^t f \|_{\C^{2B}} \leq \frac{5(d-1)}{2(d-2-\beta)}\| f
\|_{\C^{2B}} t \left( \frac{d-1-\beta}{d-1} \right)^t,
\end{equation}
for $t\geq 1$.
\finire
\section{Quantum ergodicity for equitransmitting expander graphs}
\label{sec:fuenf}

We are now able to prove theorem \ref{thm:main}. To begin with, we need a
few results on certain summations.

\begin{lemma}
  \label{lem:3}
Let $\theta\neq 1$ and $T>0$. Then
\begin{equation}
  \label{eq:52}
  \sum_{t=1}^T t \theta^t = \frac{T \theta^{T+2} + \theta - (T+1)\theta^{T+1}}
{(\theta - 1)^2},
\end{equation}
and consequently, if $|\theta|<1$,
\begin{equation}
  \label{eq:53}
  \sum_{t=1}^\infty t \theta^t = 
\frac\theta{(\theta - 1)^2}.
\end{equation}
\end{lemma}
\dimostrazione
The sum appearing in \eqref{eq:52} is of a standard type (see formula
\textbf{0.113} of \cite{gra:tis}, or \cite[p.~33]{gra:con} for a
derivation). Equation \eqref{eq:53} follows by letting $T\to\infty$.
\finire

\begin{lemma}  \label{lem:1}
  Let $T>0$ and 
  \begin{equation}
    \label{eq:40}
      \hat{w}_T(t) \coloneq \left\{
    \begin{array}{ll}
      \frac1{T} \left(  1-\frac{|t|}T \right) , & |t|<T, \\
      0, & \mbox{otherwise,} 
    \end{array}\right.
  \end{equation}
and $\theta\neq 1$. Then
\begin{equation}
  \label{eq:41}
  \sum_{t=1}^T \theta^t \hat{w}_T(t) = \frac{\theta}{T^2} \left( 
\frac{T-1+\theta^T-T\theta}{(1-\theta)^2} \right).
\end{equation}
\end{lemma}
\dimostrazione
We reverse the order of summation, to get
\begin{align}
\sum_{t=1}^T \theta^t \hat{w}_T(t) &=
\sum_{k=0}^{T-1} \theta^{T-k} \hat{w}_T(T-k),  \qquad
\mbox{\textit{via} $k=T-t$,} 
\nonumber \\
&= \frac{\theta^T}{T} \sum_{k=0}^{T-1} \theta^{-k} \left( 1- \frac{T-k}{T}
 \right)   \nonumber \\
&= \frac{\theta^T}{T^2} \sum_{k=0}^{T-1} k\theta^{-k}.
  \label{eq:43}
\end{align}
To evaluate the sum in \eqref{eq:43} we use lemma \ref{lem:3}. The result is
\begin{align}
  \sum_{t=1}^T \theta^t \hat{w}_T(t) &= \frac{\theta^T}{T^2} \left( 
\frac{(T-1)\theta^{-T-1}+\theta^{-1}-T\theta^{-T}}{(\theta^{-1}-1)^2} \right) 
\nonumber \\
 &= \frac{\theta}{T^2} \left( 
\frac{T-1+\theta^T-T\theta}{(1-\theta)^2} \right).
  \label{eq:44}
\end{align}
\finire

\dimostrazionea{theorem \ref{thm:main}}
We recall equation \eqref{eq:34} which provides the main estimate
for $V(f,B)$:
\begin{equation}
  \label{eq:34repeated}
  V(f,B) \leq \frac1{2BT}\tr \Opf^2 + \frac1{B}\sum_{t=1}^T \hat{w}_T(t)\Big(
\langle f, M^t f \rangle_{\C^{2B}} + \Ord \left( \kappa^2(d-1)^t|\curlyC_{B,2T}|
 \right)\Big).
\end{equation}
In order to estimate the error term, we have the sum
\begin{equation}
  \label{eq:47}
  \sum_{t=1}^T \hat{w}_T(t) (d-1)^t = \frac{d-1}{T^2} \left( 
\frac{T-1+(d-1)^T-T(d-1)}{(d-2)^2} \right),
\end{equation}
by lemma \ref{lem:1}.  Since $d\geq 3$, we can be sure that
\begin{equation}
  \label{eq:48}
  T-1-T(d-1) < 0,
\end{equation}
so
\begin{equation}
  \label{eq:proof_A}
  \sum_{t=1}^T \hat{w}_T(t) (d-1)^t < \frac{d-1}{(d-2)^2} 
\frac{(d-1)^T}{T^2}.
\end{equation}

The first term of \eqref{eq:34repeated} is easy to bound: since $\tr
\Opf^2 \leq 2B\kappa^2$ if $|f_b|\leq \kappa$, we have
\begin{equation}
  \label{eq:proof_B}
  \frac{\tr \Opf^2}{2BT} \leq \frac{\kappa^2}{T}.
\end{equation}

The final step needed is to bound
\begin{equation}
  \label{eq:49}
  \sum_{t=1}^T \hat{w}_T(t) \langle f, M^t f \rangle_{\C^{2B}},
\end{equation}
where, by theorem \ref{thm:general},
\begin{equation}
  \label{eq:50}
|  \langle f, M^t f \rangle_{\C^{2B}} |\leq \| f \|_{\C^{2B}}
 \| M^t f \|_{\C^{2B}} \leq K \|f\|_{\C^{2B}}^2 t 
\left( \frac{d-1-\beta}{d-1} \right)^t,
\end{equation}
for some constant $K$.  Since $\hat{w}_T(t) \leq T^{-1}$ for all
$t$, we can estimate
\begin{align}
 \sum_{t=1}^T \hat{w}_T(t) t \left( \frac{d-1-\beta}{d-1} \right)^t 
&\leq \frac1T \sum_{t=1}^\infty t \left( \frac{d-1-\beta}{d-1} \right)^t 
\nonumber \\
&= \frac1T \frac{(\frac{d-1-\beta}{d-1})}{(1-\frac{d-1-\beta}{d-1})^2}
\nonumber \\
&= \frac1T \frac{(d-1)(d-1-\beta)}{\beta^2},
  \label{eq:51}
\end{align}
making use of lemma~\ref{lem:3}. As $\| f \|_{\C^{2B}}^2 \leq
2B\kappa^2$, we end up with
\begin{equation}
  \label{eq:54}
    \sum_{t=1}^T \hat{w}_T(t) \langle f, M^t f \rangle_{\C^{2B}}
   = \Ord\!\left( \frac{B\kappa^2}{T\beta^2} \right),
\end{equation}
and combining with the other two bounds:
\begin{equation}
  \label{eq:55}
  V(f,B) = \Ord\!\left( \frac{\kappa^2}{T\beta^2} \right) + \Ord\!\left(
\frac{ \kappa^2 (d-1)^{T} |\curlyC_{B,2T}| }{BT^2} \right).
\end{equation}
\finire

\begin{remark}
  By applying the H\"older inequality we may derive the
alternative bound to \eqref{eq:54}:
\begin{equation}
  \label{eq:56}
    \sum_{t=1}^T \hat{w}_T(t) \langle f, M^t f \rangle_{\C^{2B}}
   = \Ord\!\left( \frac{B\kappa^2}{T^{1-1/q}\beta^{1+1/p}} \right),
\qquad p,q>1, \qquad \frac1p + \frac1q = 1,
\end{equation}
which may be useful to prove quantum ergodicity in cases of
families of graphs for which $\beta\to 0$ in a slow way as $B\to\infty$.
\end{remark}
\section{Examples} \label{sec:sechs}
In this section we provide two examples of families of graphs for which
our results prove quantum ergodicity when quantised with equi-transmitting
scattering matrices.
%
%
\subsection{Random regular graphs}
There exist several models for chosing a regular graph on $n$ vertices
at random \cite{wor:mor}.  We shall consider the set $\curlyG_{n,d}$
of simple, $d$-regular graphs on $n$ labelled vertices. It follows
from \cite{ben:tan,bol:app} that the size of $\curlyG_{n,d}$ obeys
\begin{equation}
  \label{eq:57}
  |\curlyG_{n,d}| \sim  \sqrt2 \rme^{(1-d^2)/4} \left( 
 \frac{d^d n^d}{\rme^d (d!)^2} \right)^{n/2}, \qquad\mbox{as $n\to\infty$.}
\end{equation}
We make $\curlyG_{n,d}$ into an ensemble of random
graphs by assigning uniform probability to each element
\cite{bol:rg1,wor:mor}. These graphs are bipartite with probability
$\littleo(1)$ as $n\to\infty$, and connected (\textit{a fortiori}
$d$-connected) with probability $1-\littleo(1)$ \cite{bol:rg2}.

In order to use theorem \ref{thm:main} to prove that quantum ergodicity
holds with probability $1-\littleo(1)$, we collect together some prior
results showing that such random graphs are expanders, and that they do
not have too many short cycles.

Random regular graphs are known to be almost Ramanujan, due to a
result of Friedmann \cite{fri:apo}, which had been conjectured by Alon
\cite{alo:eae}: for any $\epsilon>0$, with probability $1-\littleo(1)$
a random $d$-regular graph has all non-trivial eigenvalues $\mu_i$ of
its connectivity matrix bounded by $|\mu_i| \leq
2\sqrt{d-1}+\epsilon$.  This means that we can take any
$\beta<d-2\sqrt{d-1}$ in theorem \ref{thm:main}.  A weaker bound valid
for $d$ even, with a simpler proof, has been given in \cite{pud:eor},
that would also serve our purpose for those values of $d$.

For the number of short cycles in a random $d$-regular graph, such
questions have been considered in \cite{mck:sci}. For $d$ fixed, 
theorem 4 of \cite{mck:sci} reads:
\begin{theorem} \label{thm:wormald}
  Let $k=k(n)\geq 3$ satisfy $k(d-1)^{k-1}=\littleo(n)$.  Let $S=S(n)
=20Ak(d-1)^k$ with $A=A(n)>c$ for some constant $c>1$.  The probability
that the random $d$-regular graph on $n$ vertices has exactly $S$
edges which lie on cycles of length at most $k$ is less than
\begin{equation}
  \rme^{-5(d-1)^k} \left( \frac{\rme}A \right)^{S/4k}.
\end{equation}
\end{theorem}
To apply theorem \ref{thm:wormald} to our situation, let 
$k=\frac35\log_{d-1}n$,
and $S_0=\lfloor 42 n^{3/5}\log_{d-1}n\rfloor$ and define $A(n)$ by
\begin{equation}
  \label{eq:58}
  S_0 = 12 A(n) n^{3/5}\log_{d-1}n. 
\end{equation}
Then, for $n$ sufficiently large
\begin{equation}
  \label{eq:59}
  3 < A(n) \leq 3.5,
\end{equation}
and
\begin{equation}
  \label{eq:60}
  k(d-1)^{k-1} = \frac{3 \log_{d-1}n}{5(d-1)}n^{3/5} = \littleo(n),
\end{equation}
so that the conditions of theorem \ref{thm:wormald} are satisfied.  

The probability that a random $d$-regular graph has \emph{at least} $S_0$ edges
which lie on cycles of length at most $k$ is, according to theorem 
\ref{thm:wormald}, not more than
\begin{align}
\sum_{S=S_0}^\infty \rme^{-5(d-1)^k} \left( \frac{\rme}{A} \right)^{S/4k}
&\leq \rme^{-5(d-1)^k} \nonumber \\
&=\rme^{-5n^{3/5}},
  \label{eq:61}
\end{align}
for $n$ sufficiently large, since $\rme/A<1$.

Let 
\begin{equation}
  \label{eq:62}
  T = \frac{3}{10}\log_{d-1}n = \frac{3}{10}\log_{d-1}B - \frac3{10}
  \log_{d-1} \left( \frac{d}2 \right).
\end{equation}
Then \eqref{eq:61} implies 
that
\begin{equation}
  \label{eq:63}
  \mathbb{P}\left( |\curlyC_{B,2T} | \leq 42 \left( \frac{2B}d \right)^{3/5}
\log_{d-1}\left( \frac{2B}d \right)   \right) \geq 1-\rme^{-5(2B/d)^{3/5}},
\end{equation}
that is, with extremely high probability. We can then say
that
\begin{align}
 \mathbb{P}&\left( \frac{(d-1)^{T}}{BT^2} |\curlyC_{B,2T}| \to 0 \right)   
\nonumber \\
&\geq \mathbb{P} \left( \frac{(d-1)^{T}}{BT^2} |\curlyC_{B,2T}| 
\leq \frac{1400}{3}\left(\frac2d\right)^{9/10}
\frac1{\log_{d-1}(2B/d)}B^{-1/10} \right) \nonumber \\
&=\mathbb{P}\left( |\curlyC_{B,2T} | \leq 42 \left( \frac{2B}d \right)^{3/5}
\log_{d-1}\left( \frac{2B}d \right)   \right) \nonumber \\
&\geq 1-\rme^{-5(2B/d)^{3/5}}
  \label{eq:64}
\end{align}
by \eqref{eq:63}.

So with $T$ given by \eqref{eq:62}, and with probability $1-\littleo(1)$,
the right-hand side of \eqref{eq:main} converges to $0$ as $B\to\infty$,
leading to quantum ergodicity for a sequence of random $d$-regular graphs
quantised with equi-transmitting scattering matrices.


\subsection{Ramanujan graphs with large girth}  \label{sec:ramanujan}

As we have stated in section \ref{sec:notions}, the largest theoretical
value for which the parameter $\beta$ can be taken in theorem \ref{thm:main}
is $\beta=d-2\sqrt{d-1}$,  and such graphs are called Ramanujan.

Infinte families of Ramanujan graphs with $B\to\infty$ have been
constructed for certain values of $d$ only: for $d=3$ in
\cite{chi:crg}, for $d=p+1$ where $p$ is an odd prime in
\cite{lub:rg,mar:egt} and, more generally, for $d$ any prime power in
\cite{mor:eae}.  An existence proof for \emph{bipartite} Ramanujan
graphs for all values of $d$ has recently been given in
\cite{mar:ifI}.  

We will take $d=q+1$, where $q$ is a prime power.  It is known that
equi-transmitting scattering matrices of size $d\times d$ do exist.
In \cite{mor:eae} a method of construcing non-bipartite, connected,
$d$-regular,
Ramanujan graphs on $n$ vertices for a growing sequence of $n$s is given. 
Furthermore, it is proved that the girth $g(B)$ of such graphs satisfies
\begin{equation}
  \label{eq:65}
  g(B) \geq \frac23\log_{d-1}n = \frac23\log_{d-1} \left( \frac{B}{2d} \right).
\end{equation}
This girth estimate shows that these graphs are close to extremal,
since the Moore bound \cite[Ch.~23]{big:agt} gives a theoretical
upper bound of
\begin{equation}
  \label{eq:66}
  2\log_{d-1}n(1+\littleo(1)),
\end{equation}
for the girth of a $d$-regular graph on $n$ vertices.  In the case that 
$q$ is a prime, the upper bound \eqref{eq:65} has been shown to be an 
asymptotic equality \cite{big:not}.

Since the girth $g(B)\to\infty$ as $B\to\infty$, we can take any
$T<\frac12g(B)$ in theorem \ref{thm:main}, and the Ramanujan property
shows that the first term on the right-hand side of \eqref{eq:main}
can be made arbitrarily small as $B\to\infty$.   With this choice of
$T$, however, it is also clear that $|\curlyC_{B,2T}|=0$, so the
second term on the right-hand side of \eqref{eq:main} is absent,
proving quantum ergodicity for these $d$-regular Ramanujan graphs
quantised with equi-transmitting scattering matrices.

We end with two remarks concerning quantum ergodicity for equi-transmitting
Ramanujan graphs: if the bond lengths of the quantum graph are linearly
independent over $\mathbb{Q}$ then we can push $T$ up to $g(B)-\epsilon$.
Secondly, the effective logarithmic estimate for the decay of quantum
variance
\begin{equation}
  \label{eq:67}
  V(f,B)=\Ord\! \left( \frac1{\log B} \right) 
\end{equation}
is of the same order as can be rigorously proved in other systems
\cite{zel:otr,sch:ubr,deg:qva,sch:otr}, but the calculations
performed in \cite{gnu:eso} suggest that the true rate of decay should
be algebraic (which is also consistent with what is conjectured for
more general chaotic quantum systems
 \cite{eck:ate,sar:soh}).  A decay rate of $1/B$
has been proved for the quantum variance for the graphs studied in
\cite{ber:qef}, but this result aside, going rigorously beyond the
logarithmic barrier \eqref{eq:67} seems to be for quantum graphs,
as with other systems, a difficult problem.
\subsection*{Acknowledgements}
We are grateful for a number of interesting conversations regarding
this work with 
\href{http://www.math.tamu.edu/~berko/}{Gregory \name{Berkolaiko}},
\href{http://www.nottingham.ac.uk/mathematics/people/sven.gnutzmann}
{Sven \name{Gnutzmann}}, 
\href{http://math.univ-lille1.fr/~raulf/}{Nicole \name{Raulf}}, 
and
\href{http://www.maths.bris.ac.uk/~marcvs/}
{Roman \name{Schubert}}.

We thank
\href{http://www.math.u-psud.fr/~anantharaman/}{Nalini
\name{Anantharaman}} for sending us a copy of \cite{ana:qeo} prior to
publication.

We have been financially supported by \href{http://www.epsrc.ac.uk}{EPSRC}
 under grant numbers EP/H046240/1 and EP/I038217/1.

BW is grateful to the Max-Plank-Institut f\"ur Mathematik, Bonn, for
hospitality during the time that part of this work was prepared.

\appendix
\section{Matrix inequality}
Let $T\in\N$.  We recall that we defined
\begin{equation}
  \label{eq:18A}
  \hat{w}_T(t) \coloneq \left\{
    \begin{array}{ll}
      \frac1{T} \left(  1-\frac{|t|}T \right) , & |t|<T, \\
      0, & \mbox{otherwise.} 
    \end{array}\right.
\end{equation}

It is elementary to calculate the inverse Fourier transform of
$\hat{w}_T(t)$, showing that
\begin{equation}
  \label{eq:70}
  \int_{-T}^T \hat{w}_T(t)\rme^{2\pi\rmi tx}\,\rmd t = w_T(x)\coloneq
\left\{
  \begin{array}{ll}
    2\left(\frac{1-\cos Tx}{T^2x^2}\right), &x\neq 0,\\
    1, & x=0.
  \end{array}\right.
\end{equation}
For our purposes it suffices to note that $w_T$ is everywhere non-negative
and $w_T(0)=1$, for all $T$.

The following lemma is taken from \cite{sch:otr}, p.\ 1463:
\begin{lemma}  \label{lem:A}
  Let $U$ be an $N\times N$ unitary matrix, $\{u_j\}_{j=1}^N$ be an
arbitrary orthonormal basis of $U$, and $A$ be an $N\times N$ matrix.
Then if $T\in\N$, we have
\begin{equation}
  \label{eq:69}
  \frac 1N \sum_{j=1}^N |\langle u_j, Au_j\rangle_{\C^N} |^2 \leq
\frac1N \sum_{t=-T}^T \hat{w}_T(t) \tr(A^* U^t A U^{-t}).
\end{equation}
\end{lemma}
\dimostrazione
Let us denote by $\theta_j$ the eigenphases of $U$, so that
\begin{equation}
  \label{eq:71}
  U u_j = \rme^{2\pi\rmi\theta_j}u_j.
\end{equation}
By expanding the trace, we can write
\begin{align}
  \label{eq:72}
  \tr(A^* U^t A U^{-t}) &= \sum_{j=1}^N \langle u_j, A^* U^t A U^{-t} u_j 
\rangle_{\C^N} \nonumber \\
&=\sum_{j=1}^N \rme^{-2\pi\rmi\theta_j t}\langle Au_j, U^t A u_j 
\rangle_{\C^N}.
\end{align}
By inserting the representation
\begin{equation}
  Au_j = \sum_{k=1}^N \langle u_k, Au_j \rangle_{\C^N} u_k, 
\end{equation}
we get
\begin{equation}
  \label{eq:73}
  \tr(A^* U^t A U^{-t}) = \sum_{j,k=1}^N \rme^{2\pi\rmi(\theta_k-\theta_j)t}
|\langle u_k, A u_j \rangle_{\C^N}|^2.
\end{equation}
We multiply \eqref{eq:73} by $\hat{w}_T(t)$ and sum over all $t$, invoking
the Poisson summation formula to get
\begin{align}
\sum_{t=-T}^T \hat{w}_T(t) \tr(A^* U^t A U^{-t}) &=
\sum_{n=-\infty}^\infty \sum_{j,k=1}^N w_T(n+\theta_j-\theta_k)
|\langle u_k, Au_j\rangle_{\C^N} |^2 \\
&\geq w_T(0)\sum_{j=1}^N |\langle u_j, Au_j\rangle_{\C^N} |^2, 
  \label{eq:74}
\end{align}
retaining the $j=k$ and $n=0$ terms of the sums only.\finire

\end{document}